# Ethical Challenges and Evolving Strategies in the Integration of Artificial Intelligence into Clinical Practice

Ellison B. Weiner[1], Irene Dankwa-Mullan, MD, MPH[2], William A. Nelson, PhD, MDiv[3], Saeed Hassanpour, PhD[4]

[1]Dartmouth College, Hanover, NH 03755, USA

[2]Department of Health Policy and Management, Milken Institute School of Public Health, The George Washington University, Washington, DC 20052, USA

[3]Dartmouth Institute for Health Policy and Clinical Practice, Geisel School of Medicine, Dartmouth College, Hanover, NH 03755, USA

[4]Departments of Biomedical Data Science, Computer Science, and Epidemiology, Geisel School of Medicine, Dartmouth College, Hanover, NH 03755, USA

[*] Corresponding Author: Saeed Hassanpour, PhD
Postal address: One Medical Center Drive, HB 7261, Lebanon, NH 03756, USA
Phone: (603) 646-5715
Email: Saeed.Hassanpour@dartmouth.edu

# ABSTRACT

Artificial intelligence (AI) has rapidly transformed various sectors, including healthcare, where it holds the potential to revolutionize clinical practice and improve patient outcomes. However, its integration into medical settings brings significant ethical challenges that need careful consideration. This paper examines the current state of AI in healthcare, focusing on five critical ethical concerns: justice and fairness, transparency, patient consent and confidentiality, accountability, and patient-centered and equitable care. These concerns are particularly pressing as AI systems can perpetuate or even exacerbate existing biases, often resulting from non-representative datasets and opaque model development processes. The paper explores how bias, lack of transparency, and challenges in maintaining patient trust can undermine the effectiveness and fairness of AI applications in healthcare. In addition, we review existing frameworks for the regulation and deployment of AI, identifying gaps that limit the widespread adoption of these systems in a just and equitable manner. Our analysis provides recommendations to address these ethical challenges, emphasizing the need for fairness in algorithm design, transparency in model decision-making, and patient-centered approaches to consent and data privacy. By highlighting the importance of continuous ethical scrutiny and collaboration between AI developers, clinicians, and ethicists, we outline pathways for achieving more responsible and inclusive AI implementation in healthcare. These strategies, if adopted, could enhance both the clinical value of AI and the trustworthiness of AI systems among patients and healthcare professionals, ensuring that these technologies serve all populations equitably.

**Keywords:** Ethical AI in healthcare, Algorithmic bias mitigation, Patient confidentiality and consent, Transparency in clinical AI, Responsible AI deployment

# INTRODUCTION & MOTIVATION

In just the last decade, an incredible amount of progress has been made in artificial intelligence, mostly due to the advancement and application of deep neural networks, which are widely known as deep learning. The transformative impact of these novel deep learning methodologies in non-medical fields, such as robotics, autonomous driving, and speech understanding, has spurred interest and growth in the use of AI in medical applications. As a result, these models have been seriously considered for use in adjacent domains, such as biomedicine and healthcare.

To date, such promising technologies have led to the creation of sophisticated AI systems capable of performing critical clinical tasks such as medical image interpretation at the level of expert physicians.[1–4] Some of these innovative AI technologies have been created by our own team at Dartmouth.[5–12] Until recently, the creation of AI systems to assist pathologists, radiologists, and other imaging professionals required laborious "feature engineering": the manual design of algorithms to pre-process images, segment anatomic structures, detect features, and classify abnormalities. These systems often took years for AI experts to develop. Recent advances in AI have replaced feature engineering with a more efficient process based on "deep learning" from large sets of labeled or even unlabeled training data.[13] Deep learning approaches, unlike feature engineering are highly adaptable to diverse imaging tasks, allowing for more accurate, scalable and flexible solutions. These new AI systems have the potential to transform healthcare delivery, making patient care more efficient and accurate. This can reduce diagnostic errors and healthcare costs, and, critically, improve patient outcomes.

With the widespread adoption of electronic health records (EHRs) in the last decade, these records have become the most comprehensive source of clinical information for biomedical research and clinical decision-making. However, the majority of this information is embedded in free-text format in various clinical notes and reports. This unstructured free-text format, along with its associated variability and ambiguity, are major obstacles to the rapid extraction of reusable clinical information from medical records and its subsequent use for biomedical research and clinical decision-making. Recent advancements in natural language processing (NLP), fueled by the exceptional performance of large language models (LLMs) across diverse tasks, have opened new avenues for AI models to bring innovations in this area.[14–21] These NLP methodologies can address the critical challenges of information extraction from unstructured data and build informatics methods to unlock this information for translational research and clinical care.[22–24] These novel natural language processing and machine learning methods can enable the distillation of critical insights from heterogeneous and complex free-text medical records, such as clinical notes, radiology reports, and pathology reports, to provide intelligent tools for translational research and clinical decision support. The integration of NLP into clinical environments can be a pivotal advancement in healthcare, offering the potential to unlock profound insights from EHRs and significantly enhance patient care.

Undoubtedly, the field of medicine is starting to quickly undergo a significant transformation by the integration of AI and machine learning and will continue to advance at an unparalleled rate. As these changes unfold, the medical community, including health care organizations and clinicians, faces major concerns such as ensuring the ethical implementation of AI while legislation lags. The field of AI ethics considers the responsible development, deployment, and use of AI against a backdrop of business-driven innovation that considers the current legal and ethical standards within the industry. The rapid pace of development of AI demands an inclusive conversation its ethical use among experts, which has prompted a plethora of research. Neither should one group of people solve this complex issue, nor should the forum become an echo chamber. Accordingly, the following paper presents widely discussed challenges and proposed directions to solve challenges around ethical use of AI in clinical practice. We will review the merits of these recommendations and the aspects that need to be further explored in this domain.

# CORE ETHICAL CHALLENGES

In this section, we will present major ethical concerns associated with the integration of AI in clinical practice.

## 1. JUSTICE AND FAIRNESS

    **A. Eliminate embedded bias in algorithms to ensure current bias is not exacerbated.**

        i. Justice and fairness are core ethical principles in healthcare AI, drawing on "distributive justice"—the equitable distribution of resources to prevent disparities—and "procedural justice", which ensures fairness in decision-making processes. In clinical settings, these principles demand that AI systems not perpetuate existing biases or inadvertently favor certain patient groups over others. Fairness in AI is particularly critical given the potential of biased algorithms, developed on non-representative data, to lead to unequal access, lower quality of care, and even misdiagnosis among marginalized populations.

        ii. In a literature review of 45 sources, justice and fairness was the ethical issue of highest concern in 24 of the articles, arising in concert with themes such as bias, discrimination and equality.[25] Strategies noted amongst the literature took both an algorithmic and data perspective, suggesting that developers purify algorithms of decision support tools, manage fairness constraints and distribution, guarantee responsible data collection, and encourage the cooperation of stakeholders in AI development.

            1. A widely used healthcare algorithm calculating overall health status (as measured by number of chronic health conditions) found that black and white patients were assigned the same level of risk, despite black patients being much sicker than their white counterparts. Health care costs were used as a proxy for the need for medical attention. Thus, the algorithm included implicit racial bias as less money is typically spent on black versus white patients. If this disparity was accounted for, black patients would receive 46.5% additional care, as opposed to an initial 17.7%.[26]

## 2. TRANSPARENCY

    **A. Address the problem of explainability within AI models by providing the ability to verify results and in-program qualities.**

i. Transparency is a foundational principle of AI ethics, particularly relevant in healthcare, where trust and accountability are paramount. Within this context, transparency involves multiple dimensions: "data transparency" (clarity on the data sources and how representative they are), "algorithmic transparency" (insights into the underlying model structure and assumptions), "process transparency" (disclosing the development steps, including human interventions and choices made during training), and "outcome transparency" (communicating how the system generates its results). Closely linked to transparency are "explainability"—the ability to clearly describe how an AI model arrives at specific conclusions or predictions—and "interpretability", which is the extent to which humans can understand the cause-and-effect relationships within a model's functioning. Together, these aspects help address the "black-box" nature of AI, allowing healthcare professionals and patients to trust and comprehend AI-driven decisions and thereby supporting informed, ethically sound clinical practices.

ii. The "black-box" problem of AI gives rise to the crisis of interpretability, which becomes even more significant in healthcare when solutions must be explained to caregivers and patients alike. Current explanations given by AI are often inaccurate and sometimes misleading, given that they are made post-hoc (i.e., after results are generated). AI developers, health care organizations, and clinicians must mediate this problem of explainability by ensuring the tools are trustworthy, user-friendly, and human-centric.[27]

iii. Patients must know and understand the process behind healthcare decisions made by AI, and require care givers to explain to patients the limitations and reason for such AI-driven decisions.[28]

iv. Healthcare providers also face uncertainty in decision making due to the "black-box" issue of transparency, furthering the significance of creating explainable AI.

**B. Trustworthiness: Establish representative data to train and test an AI model while ensuring transparency.**[27]

i. Trustworthy AI systems in healthcare are grounded in transparency, explainability, and interpretability, as these qualities collectively ensure that models are safe, reliable, and fair across diverse patient populations. For healthcare providers to trust AI-based insights, they must be confident that these insights are generated from "representative data" that covers relevant demographic and clinical variations. Transparency plays a critical role in building this trust, enabling clinicians and stakeholders to understand not only the sources and representativeness of data but also the specific processes and algorithms involved. By ensuring transparency in all stages—from data collection and preprocessing to model design and training—developers provide the groundwork for interpretable, explainable systems that foster trust and improve clinical decision-making reliability.

ii. Machine learning is trapped in a circular paradox: AI applications have a "large need for data to keep learning and improving, hence becoming safer, hence more trustworthy".[29] For models to be considered trustworthy, they must be well-trained with a representative dataset. Rare health conditions lack extensive amounts of data, thereby making the software less trustworthy in its results and explanations. However, big data can be worrisome in terms of reliability and complexity, thus reducing the model's trustworthiness and, by extension, transparency.

## 3. PATIENT CONSENT AND CONFIDENTIALITY

**A. Acquire informed consent to use patient data and ensure anonymity.**

i. Patient consent and confidentiality are central ethical concerns in healthcare, especially as AI increasingly utilizes vast amounts of patient data. "Patient consent" ensures that individuals have control over how their personal health information is used, aligning with the ethical principle of autonomy. "Confidentiality" protects this information from unauthorized access or disclosure, fostering trust between patients and healthcare providers. In the context of AI, these principles face unique challenges: AI models often require large, diverse datasets, which can sometimes lead to conflicts between the need for comprehensive data and patients' rights to privacy. Without robust consent mechanisms and stringent confidentiality protections, the use of patient data for AI can risk violating patient privacy, undermining trust, and eroding the ethical foundation of healthcare systems. Addressing these challenges is therefore essential to ethically integrating AI in clinical practice.

ii. An inherent point of conflict exists in the decision of whether to prioritize comprehensive datasets for AI models, or guaranteeing consent of patients even if confidentiality is already ensured. Of course, data leaks are still possible thus preserving the importance of informed consent. In relation to issues of trustworthiness and transparency, increasing the rate of consent is an important area for improvment in the field.

**B. Respect the privacy rights of users and third parties.**

i. Patient trust and autonomy is affected by their right to privacy. Most "mobile disorder detection systems" risk data hacking as they use mobile devices to acquire signals, transfer, analyze, and forward the results to users in a stored database.[25] The same is true for online systems.

**C. Remind patients of their ability to opt out at any time and empower them to exercise autonomy in choice.**

i. To respect patient autonomy, experts suggest clinicians discuss the topic in terms of trust, shared decision-making and legal responsibilities of clinicians; a uniform understanding of issues involved in patient-clinician relationships; and ensure transparency regarding the use of AI to convey to patients that human-judgment takes priority over AI systems.[25]

ii.  While consent may be acquired at the time of use, many AI models use this data in an ongoing fashion to continue updating. Thus, an additional challenge arises in considering whether patient consent should be frequently in discussion. Patients should have the right to opt out at any time, but this may impact the model, which has already applied the data to learning algorithms.

## 4. ACCOUNTABILITY

### A. Properly delegate and accept responsibility for transparent and ethical conduct.

i.  Accountability is a critical ethical concern in healthcare AI because it establishes clear responsibility for the outcomes and impacts of AI-driven decisions on patient care. In traditional medical practice, accountability is well-defined, with clinicians held responsible for diagnosis and treatment decisions. However, AI systems introduce complex layers of responsibility, involving model developers, healthcare providers, and institutions, all of whom play a role in implementing and overseeing AI technologies. When AI systems produce errors or unsafe recommendations, determining accountability becomes challenging, especially if the system's inner workings are opaque or lack clear documentation. Without well-defined accountability, patients and providers may have limited recourse in cases of harm, leading to potential safety risks and diminished trust in AI applications. Thus, robust accountability frameworks are essential to ensure that all stakeholders involved in healthcare AI—from development to deployment—are aligned in prioritizing patient safety and ethical conduct.

ii.  Although AI often commits errors, the stakes are much higher when considering a patient's health. The possibility for AI to provide unsafe advice and treatment protocol is not negligible, and we must be able to attribute the harm to a specific source. This interactions between AI model developers, organizational leaders, and health care providers can create risk in that they may not be inclined to take responsibility for errors. AI developers may fear monetary consequences over ethical considerations, while medical professionals may inadvertently place patients more at risk due to a subconscious feeling of immunity from the AI system. There is a clear misalignment of risk and return that requires ethical consideration and emphasis.

iii.  Conflicting advice given by AI and medical experts will prove difficult for healthcare providers, as models often lack a measure of certainty.

iv.  The role of organizations and healthcare institutions in terms of monitoring AI, assessing AI implementation, and oversight need to be included in this measure of accountability. Organizations promote certain regulations under which developers and providers are bound by. Another challenge relates to whether groups must disclose if they are using AI as part of shared decision making.

## 5. PATIENT-CENTERED AND EQUITABLE CARE

**A. Ensure AI complements the role of primary caregivers .**

i. AI systems should be designed to enhance, not replace, the primary caregiver's role in building trust and rapport with patients. The therapeutic relationship between healthcare providers and patients is foundational to effective care, ensuring patients feel understood, valued, and supported. AI should be integrated in a way that maintains and strengthens this essential human connection.

ii. AI can provide real-time insights, streamline diagnostic processes, and offer data-driven recommendations that assist caregivers in critical, complex situations.[27] By enhancing diagnostic accuracy and efficiency, AI allows caregivers to focus on the personal and emotional aspects of patient care, where human empathy and judgment are irreplaceable.

iii. To align with the goals of patient-centered care, AI tools must be adaptable, ensuring recommendations are consistent with individual patient needs and preferences. This requires transparency in how AI systems generate insights, allowing clinicians to interpret AI outputs in a way that respects each patient's unique context and circumstances.

iv. AI must be deployed within clear safety guidelines, with clinicians maintaining final responsibility for treatment decisions. AI outputs should be treated as supportive tools rather than definitive instructions, empowering caregivers to make informed, context-aware choices that best serve their patients.

**B. AI should communicate with empathy and equity to patients.**

i. AI systems must be developed to support equitable healthcare delivery, which requires careful attention to socioeconomic, gender, and ethnic factors that can affect patient care. Research has shown that disparities, such as black women receiving less care when reporting pain,[30] highlight the potential for biases in healthcare that AI systems could inadvertently perpetuate if not properly designed.

ii. To avoid exacerbating these biases, AI models should be rigorously tested across diverse demographic groups to ensure fair and equitable outcomes. This includes utilizing diverse datasets and bias-detection methods to identify and address potential disparities in AI predictions or recommendations. Ensuring that AI is trained on representative data can help minimize biased responses and improve fairness across patient populations.

iii. AI systems must also be designed to communicate in a way that respects patients' emotional and psychological needs. This requires embedding a sense of empathy within AI interactions to create responses that are human-centered and sensitive to patients' vulnerabilities. The goal is for AI to offer support in a manner that feels respectful and considerate, rather than impersonal or dismissive, fostering patient comfort and trust.

iv. Finally, AI-driven insights and recommendations should provide clear, understandable explanations that help patients feel informed and reassured. By delivering transparent and granular explanations of diagnoses or treatment plans, AI can empower patients to make informed healthcare decisions and feel more involved in their care. This human-centric approach reinforces empathy and equity, helping to mitigate biases and promote fairness in healthcare.

# EMERGING IDEAS

In the previous section, we covered the current concerns regarding the use of AI in healthcare applications. Now we will discuss the potential frameworks that have been proposed to address these concerns and ensure ethical use of AI.

## 1. MULTI-SCALE ETHICS

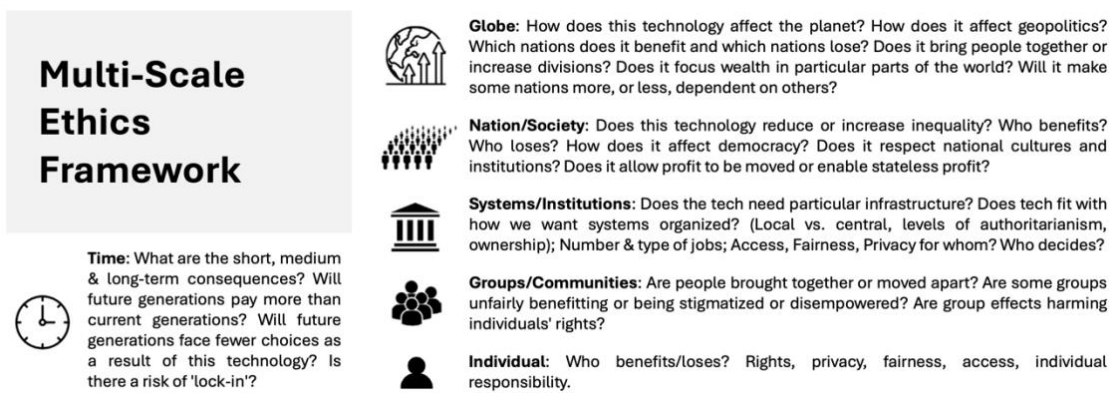

**Figure 1.** Multi-Scale Ethics Framework proposed to evaluate the ethical issues of AI at interactive levels of community. Reprinted from "Multi Scale Ethics—Why We Need to Consider the Ethics of AI in Healthcare at Different Scales," M. Smallman, 2022. *Science and Engineering Ethics*, 28(6), 63.

This framework outlines AI as a "socio-technical" system, which demands we consider social and ethical implications resulting from the use of technology in healthcare.[31] Smallman identifies the gap in our current approach as the lack of a framework contextualizing the risks and benefits of AI. Currently, most research focuses on the ethical threats facing an individual; for instance, privacy, autonomy, and transparency, to name a few. However, there are unique threats facing each "level" and over different time scales, which we should not consider solely as an aggregate of individual effects (Figure 1). Smallman anticipates that this framework will help us determine the "pattern" of risks and benefits, anticipate where they may continue to occur, and use this information to decide how to implement the technology. This framework also brings into question the role of patients in community dialogues. Given that individuals comprise a basic level of knowledge, they must be included in decision making around concerns of ethics. Dialogues with patients and experts alike are likely to yield the most comprehensive and stable solutions to current challenges. This make take the form of public forums and panels.

## 2. "SHIFT" ACRONYM FOR STANDARDIZATION

**A. Sustainability, Human Centeredness, Inclusiveness, Fairness, Transparency (SHIFT)**

i. In a thematic analysis of recent literature,[32] the following subthemes were noted with the corresponding frequency (of 253 articles): responsible local leadership (14); social sustainability (22); embedding humanness in AI to meet ethics of care (20); the role of health professionals in maintaining public trust (32); developing artificial wisdom through interdisciplinary collaboration (6); inclusive communication and involvement in AI governance (19); alleviating algorithmic and data bias (89); data representation and equality (22); health disparity in low resource settings (22); safeguarding personal privacy (54); explainability of AI-driven models and decisions (56); addressing the loss of confidentiality by legislation (16); user empowerment (6); and informed consent for data use (58). Figure 2 illustrates the distribution of key subthemes in AI ethics literature from this review, highlighting the most prevalent concerns in responsible AI implementation in healthcare.

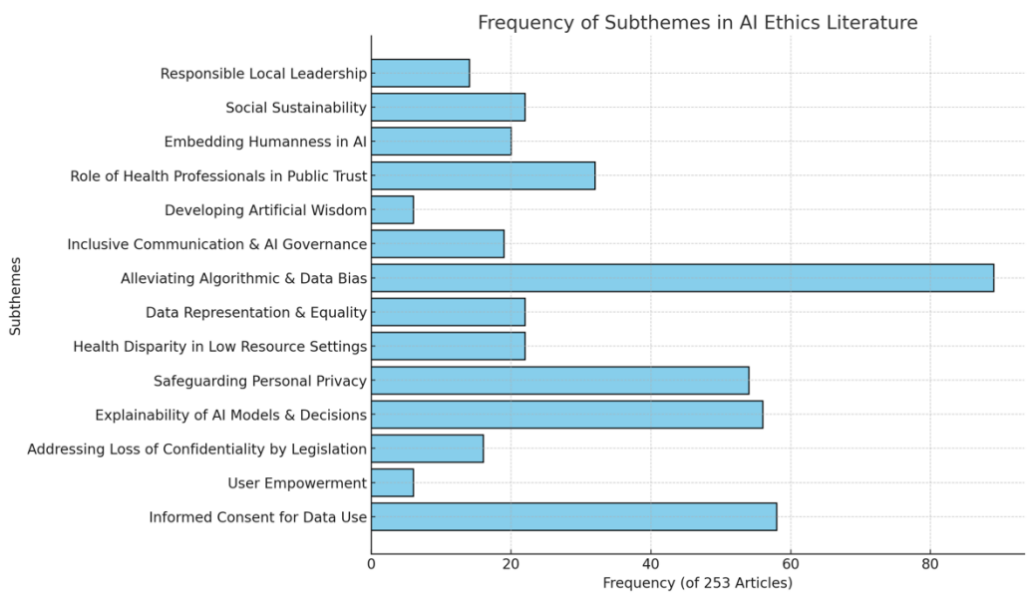

**Figure 2.** Frequency of key subthemes in AI ethics literature. Algorithmic bias, informed consent, explainability, and privacy emerge as the most prevalent concerns in responsible AI implementation in healthcare.

Using standardized acronyms like "SHIFT" may help the field come to a consensus on the challenges that should be prioritized and the initiatives that we should take to protect patients and communities. This work is important to educate those involved in the variety of applications of AI in medicine may. Siala and Wang review examples of responsible initiatives such as: connecting algorithm outputs directly to the human-decision making process in the medical field; implementing a centralized institutional review board; and aggregating different types of patient data from health information systems to increase explainability, to name a few.[32]

## 3. RESPONSIBLE INNOVATION FOCUSED ON INCLUSION

Responsible innovation in healthcare AI involves developing technologies that are ethical, equitable, and designed with the needs of all patient populations in mind. It emphasizes creating AI systems that not only achieve technical objectives but also actively prevent harm, mitigate bias, and promote inclusivity. The concern of bias embedded within algorithms, which can exacerbate existing inequities in healthcare, remains one of the most pressing issues in the field. Julia Trabulsi, a BioTech product lead and advisor, recommends that developers build controls that matter, consider everyone, and put people before profits to ensure responsible innovation. In a guest lecture for a "Neuromarketing and Consumer Neuroscience" course at Dartmouth College on February 7, 2024, she emphasized the importance of oversampling underrepresented communities to balance data and create an even distribution for various demographic groups, which helps reduce bias. Trabulsi acknowledged that data will inherently reflect societal biases, so prioritizing inclusivity and fairness at each stage of AI development should be of utmost importance to achieve responsible, ethical AI in healthcare.

## 4. ALGORITHMOVIGILANCE

**A. This term, inspired by "pharmacovigilance", demands consistent evaluation of algorithms to mitigate bias and ensure fairness.**

The process of eliminating bias "requires vigilance at every stage of an AI system's development" due to the common tendency of developers to unconsciously introduce bias due to sample size, historical bias, representation bias, sponsorship bias (particularly regarding the integrity of clinical studies) and self-serving bias.[27] Polevikov proposes the following best practices when collecting and analyzing data: incorporating effect size and confidence intervals, utilizing appropriate sample sizes, avoiding "data shopping" and self-serving biases by adopting transparency in data analysis.

At the forefront of our concerns is ensuring the use of AI does not disrupt healthcare delivery and the provider/patient relationship, but rather enhances it. A potential solution could be to include a measure of uncertainty in the model to aid decision making, as providers can get a sense of the reliability of the advice provided.

In the following Discussion Section, we will cover important takeaways from the current state of the field and the directions that need to be explored to ensure the use of AI in the medical field aligns with ethical standards.

# DISCUSSION

## IS THE DEVELOPMENT OF HEALTHCARE AI FAIR AND NOT BIASED?

Considering the rapid pace of development in the field of medicine, it is safe to assume that the integration of AI and machine learning in the healthcare sector will continue to grow at an unprecedented rate. Thus, staying ahead of the curve in terms of ethical standards and legislation has become very challenging. Current work in this domain highlights issues such as justice and fairness, transparency, patient consent and confidentiality, accountability, and patient-centered and equitable care . These concerns are valid as understanding of these models and their implementation is largely acquired through implementation itself. Accordingly, the development of healthcare AI is not consistently fair and unbiased, but commitment to this discussion has the potential to eliminate unfair biases in the system. Clearly, not all concerns are uniform, given that the issues are demanding and complex. The current direction of the healthcare sector demands medical students be prepared to manage AI beyond being passive users of information technology. Furthermore, they must apply their understanding of the "patient-first" ideology in the context of artificial intelligence by ensuring autonomy, privacy, transparency, fairness and accountability.

Many agree that the concern of bias is the most pressing when considering the integration of AI into the healthcare field, especially considering the lack of industry standardization in regulation and review. When regulatory bodies conduct bias audits, they are typically only completed during the phases of model development and validation, which exposes a large portion of development to bias. However, even AI imaging products cleared by the FDA under further review revealed problematic practices. Only 64% of products from November 2021 used clinical data for validation, while the data lacked information on demographic and technical details. Specifically, only 4% provided patient demographic and 5% provided machine specification. Current frameworks are lacking in standard, as only 34% were validated by multiple institutions and reported said institution.[33]

Complications with bias often arise when there is a mismatch between the training population of an algorithm and the real-world data it encounters in clinical settings. For instance, researchers have found that Blacks, Hispanics, or females are less likely to receive cardiopulmonary resuscitation (CPR), regardless of income level or location where the cardiac arrest occurs.[34,35] As a result, these patients are often underrepresented in cardiac imaging cohorts, which impacts the accuracy of machine learning models trained on such data for predicting cardiac disease outcomes. This exclusion can have numerous downstream effects, as clinical parameters and biases in care influence the radiology datasets used for machine learning, which ultimately affect ordering patterns and model performance. NIH-funded initiatives like AIM-AHEAD and Bridge2AI identified this misalignment between representative training data and clinical diversity as one of the most difficult challenges in mitigating bias in existing datasets.[33] Some experts suggest using datasheets, or checklists, to help developers ensure that datasets are representative and balanced, reducing potential biases and improving fairness in AI models.

Bias may interfere in various stages of model development through mechanisms such as exclusion, annotator, funding, and objective mismatch bias, to name a few. Many attribute flaws in the review system to the lack of diversity and representation on development teams. The AI field would benefit from encouraging clinicians, analysts and patient advocacy groups to exchange ideas and debate.[33]

One potential mechanism to address the ethical application of AI in healthcare is to require developers to submit their programs to an oversight review mechanism before use in the healthcare sector, in which experts from many disciplines collaborate in their vigilant search for bias, empathy, and transparency. This process could serve as a critical checkpoint to ensure that AI systems align with ethical standards and are robust against potential harms. Furthermore, the concern of exacerbating bias remains the most pressing issue acknowledged by experts.[32] Thus, we must utilize the full force of communication to reach disparate groups to create representative models, and in discussions among experts to verify the mitigation of bias, as well as implement appropriate sampling methods of patient populations.

## IS THE DEPLOYMENT OF HEALTHCARE AI PATIENT-CENTERED?

The American Medical Association (AMA), as recently as 2023, committed to develop policy on potential unforeseen conflicts created by the use of AI in healthcare and notes the widely accepted concerns aforementioned.[36] However, the current field of AI ethics lacks a standardized protocol and consideration of the unique effects at each level of society. The AMA's commitment is a step in the right direction, although, Smallman's "Multi-Scale Ethics" model crucially demands a broader understanding of ethical concerns that will inform our approach to eliminating bias and implementing ethical use of AI. Further, protocol for responsible innovation and extreme vigilance in monitoring algorithmic biases will only increase the ethical use of AI in healthcare. This includes the practice of balancing data by overrepresenting certain communities to mitigate bias in developing and training AI models.

Current regulatory frameworks are constantly evolving given the nature of the field but should be designed to address topics such as mandated disclosure of training data, standards for accuracy, reliability, and consistency, problems of liability, ethical and moral use guidelines, as well as measures that must be taken amongst developers to ensure privacy, continuous oversight, monitoring.

Two relevant models that effectively balance concerns of inspiring innovation while considering safety and privacy include the European Union's General Data Protection Regulation (GDPR) for data protection and the US Food and Drug Administration's (FDA) regulation of AI-based medical devices.[37] The former utilizes a risk-based approach to monitor AI by factoring intended use and potential harms into the framework. Uses of AI like social scoring and mass surveillance are classified as unacceptable, high, and limited – high risk includes the use of AI in areas like critical infrastructure, medical devices, and employment. The FDA conducts a pre-market evaluation for devices deemed higher-risk, ongoing post-market monitoring and stringent manufacturing and quality control standards. The Federal Trade Commission (FTC), in a joint statement with other U.S. federal agencies, announced that it has required firms to destroy algorithms trained on inappropriately collected data.[38] The FTC also regulates AI that aids socioeconomic decisions, monitors bias and privacy violations, as well as marketing claims related to AI healthcare products with the intention of preventing deceptive and biased practices.[39] The Equal Employment Opportunity Commission said in the same statement that it claims responsibility for enforcement over discrimination in AI and automated systems. The National Institute of Standards and Technology (NIST) is developing a US AI Bill of Rights covering the following five areas: safety and efficacy, algorithmic discrimination protections, data privacy, notice and explanation (to ensure informed consent), and human alternatives, consideration, and fallback. The US AI Executive Order on Safe, Secure, and Trustworthy Artificial Intelligence relies on core principles of fairness, transparency, safety, privacy protection, inclusiveness, sustainability, reliability, and accountability.

Most frameworks are post-hoc by nature, entailing a resubmission process which disincentivizes thorough review of the ethical status of an algorithm at first.[33] Moreover, current evaluation strategies have high costs of validation due to factors such as standards for the preparation and development of a data set, a diverse team of experts in both machine learning and in the clinical setting, and the burden of receiving regulatory approval. While data regulations rightfully limit access, they also constrain available data. This renders responsible innovation and model development difficult despite protecting patient confidentiality.

## IS THE USE OF HEALTHCARE AI ETHICAL? IF NOT, HOW CAN POLICY BE CHANGED TO ENSURE ETHICAL IMPLEMENTATION?

All current and promising methods, however, fail to ensure that the integration of AI will be seamless and consistently ethical. Particularly, one shortcoming that is not addressed by any of the previously mentioned directions is the collaboration between disciplines. Often, there is a perception that ethics—in the context of healthcare, business, or any other professional field—simply involves consulting a book or a set of published guidelines, and the ethics "work" is done. This should not be the accepted view of ethics for any organization seeking to utilize AI in healthcare. Instead, ethics should be seen as an essential and continuous process of internalizing moral decision-making. This means that organizations need to engage regularly in multidisciplinary discussions about the ethical application of AI. Such a process should clearly involve the presence of trained healthcare ethicists.

Much of this conversation of AI ethics occurs at a level entirely separate from the practice of the field. In other words, there is a disconnect between "thinkers" and "doers" as the heavy load of engineering and development responsibilities in the field prevents them from engaging with protocols about the ethical use of AI. Considering the lack of standardized guidelines, organized vigilance, and the dynamic nature of the field, there should be new initiatives to connect ethicists, AI developers, researchers, healthcare practitioners and leaders to jointly think about and communicate the benefits and common pitfalls of using AI in healthcare and medicine. As mentioned previously, these community dialogues should include patients alongside experts as many of the challenges in play revolve around protecting patients. Moreover, diversity should not only be represented in the development of programs, but also in research teams, institutional advisory committees, and leadership positions. Even further, establishing partnerships with minority-serving institutions, community organizations, and other stakeholders will yield a more innovative and inclusive research environment. The use of artificial intelligence has a unique and unknown impact on healthcare which demands an ongoing consideration of ethical issues. To remain vigilant and open-minded, we must acknowledge the existence of important issues that we may have missed as well as continuing to study the development of the field via research and discussion.

At present, there is a strong need to study the long-term effects of the use of AI in the medical field, specifically in areas like patient outcomes and overall efficiency to consider how to adopt best practices moving forward. Studies on treatment efficacy, cost-effectiveness, patient satisfaction, and the impact on healthcare workflow are necessary for an evolving field.[37] These proposed research inquiries should also prioritize an interdisciplinary and collaborative approach amongst developers, clinicians and healthcare providers, ethicists and policy makers to gather diverse and realistic perspectives on the ethical use of AI. In anticipation of new technology and devices powered by AI that will emerge in the coming years, we must acknowledge that the ethical concerns associated with these products will also be unique. Regulatory bodies comprised of a diverse staff must consistently seek feedback and criticism to remain current and anticipate future challenges. As expressed by Siala and Wang under the aforementioned SHIFT framework, future studies should not limit their focus to ethical considerations, but also anticipate the social implications of scientific analysis and evolution.[32] However, the integration of AI into the medical field should remain optimistic. Healthcare professionals will be relieved of certain repetitive tasks, allowing them to redirect their attention towards patients.[37] Patient-centric care will be revitalized, as will the accuracy of diagnostic and therapeutic decisions with machine learning algorithms. Yet, human judgment and expert opinion will not be threatened by the use of AI, but rather supplemented and improved as provider-patient interactions are critical.